# Tunable Memory and Activity of Quincke Particles in Micellar Fluid

Yang Yang[a], Meng Fei Zhang[a], Lailai Zhu[b], Tian Hui Zhang[a*]

[a] Center for Soft Condensed Matter Physics and Interdisciplinary Research &School of Physical Science and Technology, Soochow University, Suzhou, 215006, P. R. China
[b] Department of Mechanical Engineering, National University of Singapore, 117575, Singapore

Memory can remarkably modify the collective behaviors of active particles. We show that in a micellar fluid, Quincke particles driven by a square-wave electric field exhibit a frequency-dependent memory. Upon increasing the frequency, a memory of directions emerges whereas the activity of particles decreases. As the activity is dominated by interaction, Quincke particles aggregate and form dense clusters in which the memory of the direction is further enhanced due to the stronger electric interactions. The density-dependent memory and activity result in dynamic heterogeneity in flocking and offer new opportunity for study of collective motions.

## Introduction

In active systems, energy is supplied and consumed to propel the motion of constituent particles.(1) Various collective motions can emerge spontaneously in active systems.(2) To describe the collective behaviors, several theoretical models have been developed.(3) Among them, active Brownian particles (ABP)(4) and Vicsek-type models (VM) with alignment(2, 5, 6) are widely employed to interpret the emergence of collective motions. In these models, the dynamics of active particles is assumed overdamped and inertia is neglected. Theoretically, however, inertia-induced memory can remarkably modify the dynamics of active systems and results in complicated collective behaviors. For example, as Vicsek-type models are coupled with a inertial-induced memory in the form underdamped angular dynamics, a rich variety of collective motions, including vortex lattice and active foams, emerge.(7) In overdamped active nematics, friction can create a strong memory of local deformation, and then affects the dynamic behaviors of defects.(8) In viscoelastic fluids, the stress-induced memory produces a phase difference between the propulsion force and the velocity of active particles, giving rise to a persistent circular motion.(9)

Dielectric particles dispersed in a weakly conducting liquid can be driven by an electric field to roll via the mechanism of Quincke rotation.(10, 11) Quincke particles sitting on a solid surface serve an experimental model of active colloids which can form large-scale collective motions.(12) The torque of Quincke rotation arises from the interaction between the dipole moment $\vec{P}$ and the electric field $\vec{E}$. For particles driven by a constant $\vec{E}$, the direction of rolling is determined by the fluctuation-induced in-plane component of $\vec{P}_\perp$, which is maintained in the following movement. As Quincke particles are subjected to a square-wave electric field which are periodically stopped and reset, their behavior becomes complicated.(13-15) It was argued that the complex dynamics of Quincke particles arises from the competing between the relaxation of polarization and the reset of Quincke rotation. The incomplete relaxation of polarization remains the direction of propulsion, giving rise to a memory of propulsion. As the stop-time of electric field is longer than the depolarization time, the memory will not occur.(13) So far, the inertia of Quincke particles is neglected in the studies of collective motions.(14-18)

However, in this study, we show that in a micellar fluid, the dynamics of Quincke particles driven by a square-wave electric (SWE) exhibits a strong signature of inertia-like behavior: a finite acceleration time is necessary for particles to reach their equilibrium speed. Also, there is a deceleration time for a moving particle to relax its momentum as the electric field $E$ is switched off. As the half period of $E$ is less than the deceleration time, the direction of velocity will be retained, giving rise to a memory of the direction of motion. As the half period of electric field is less than the relaxation time of charge, the coupled directions of propulsion and velocity are retained simultaneously, giving rise to a persistent directional motion. Nevertheless, the speed in persistent directional motions under SWE is not a constant but oscillate periodically with $E$. The resulting effective activity in terms of the mean speed is then frequency-dependent and much smaller than that observed under a constant electric field. Most importantly, it is found that the memory of direction can be significantly enhanced in dense clusters as the electrostatic interactions between particles are strong.

## Experimental Methods

Polystyrene particles of a diameter $10.0\ \mu m$ are dispersed in the mixture of AOT/hexadecane $(0.15\ \mathrm{mol/L})$. The suspension is sealed in a cell constructed by two ITO-coated glass slides. All particles settle down on the bottom electrode. As a uniform electric field $\vec{E}$ is applied, the particles become polarized. In this system, the charge relaxation time of solid particles is larger than that of the solvent such that the dipole in the particle is antiparallel to $\vec{E}$ (Fig.1a).(12, 19) As the dipole moment $\vec{P}$ is disturbed away from its equilibrium state by fluctuations, an in-plane component $\vec{P}_\perp$ emerges and a destabilizing electric torque $\vec{P} \times \vec{E}$ forms, which drives the particle to rotate (Fig.1a). However, due to the stabilizing effect of the viscous fluid, the amplitude of $\vec{E}$ has to exceed a critical value $E_c$ to initiate the rotation.(10, 11) In a steady state, Quincke particles run at an equilibrium translational speed $v$, which scales with $E$ by $v \sim \sqrt{(E/E_c)^2 - 1}$.(11) The trajectories of particles in this study are recorded with a resolution of 1.0μm/pixel and a rate of 1000 frames per second. The positions of particles are located by IDL routines with an accuracy of 0.1pixel.(20) Therefore, the speed of particles is measured with an accuracy of 0.1μm/ms.

## Results and discussion

In this study, a square-wave electric field (SWE) of period $T$, which periodically hops between a peak value $E_p (> E_c)$ and a ground value $E_g (< E_c)$, is employed to propel the particles. Observations are conducted first with $E_g = 0$ (Fig.1b up). In this study, the durations of $E_p$ and $E_g$ in each cycle are equally set as $T/2$. Accompanying the oscillation of $E$, Quincke particles oscillate periodically between an active state ($v > 0$) and a rests state ($v = 0$) (Fig.1b bottom). However, Fig.1c shows that there is an acceleration process before the particles reach the equilibrium speed. As $E$ is switched off, there is also a deceleration process by which the particle relaxes to rest smoothly. To quantify the time $\tau_a$ for acceleration and the time $\tau_d$ for deceleration, the speed is averaged over thousands of cycles (SFig.1a). $\tau_a$ and $\tau_d$ are then estimated as ~13ms and ~ 4ms respectively.

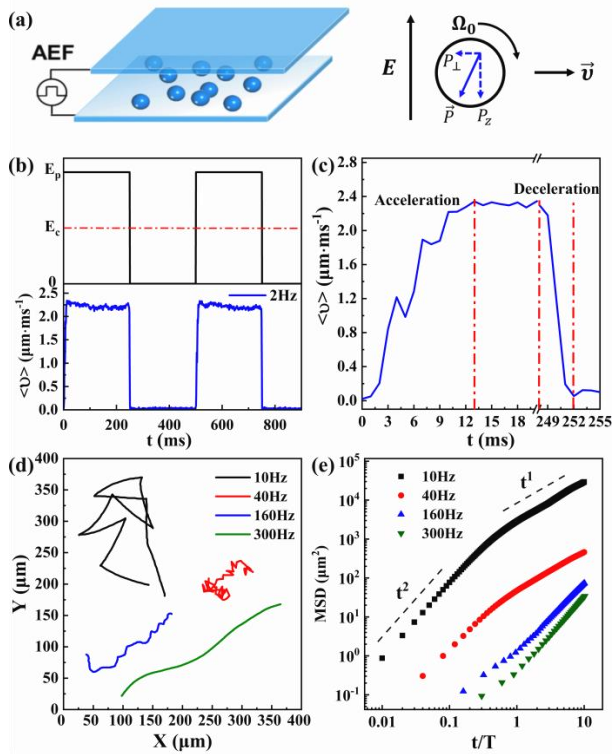

**Figure 1** Properties of Quincke particles under a square-wave electric field. $E_p = 2.1 E_c$ and $E_g = 0$. (a) Schematics of the experimental setup. (b) Square-wave electric field and the oscillating speed. (c) Evolution of speed in one cycle at 2Hz. (d) Trajectories of particles at different frequencies. (e) Mean square displacements (MSD) observed at different frequencies.

Both the acceleration and the deceleration suggest that there is a strong inertia-like feature in the dynamics of Quincke particles. However, as a colloidal system, Quincke systems are characterized by a very small Reynolds number.(12, 21) Here, in this study, Reynolds number $R_e = \frac{vR\rho_f}{\eta} \approx 9.0 \times 10^{-4}$, where the density $\rho_f$ of the solvent is $0.77 \, g \, cm^{-3}$, the typical velocity of particles $v$ is on the order of $\mu m/ms$ and the viscosity $\eta$ of the solvent is $4.3 \times 10^{-3} \, P_a \, s$.(22) Therefore, the dynamics of Quincke particle is overdamped and the inertia of particle is neglected.(13) Nevertheless, the concentration of AOT used in this study as well as in previous studies(12, 18, 19) is far larger than the critical micellar concentration (1mM) such that the fluid is filled with nanometer-sized micelles of AOT.(23) The nanometer-sized micelles can significantly modify the conductivity and the viscosity of the fluid.(24) In micellar fluids, overlapped micelles form an extensive elastic network and the viscoelasticity of the fluid becomes significant. In the viscoelastic fluid, the motion of particles deforms the local microstructure of the micelle network. The resulted stress exerts a friction on the motion of particles. The magnitudes of the deformation and the stress are strongly dependent on the velocity of particles (shear rate), giving rise to a velocity-dependent friction coefficient and nonlinear viscoelasticity.(25) Moreover, the non-instantaneous response of micelle network and the lagged relaxation of stress results in a memory on the motion of driven particles.(9, 26, 27) The memory about the motion in the viscoelastic fluids leads to a time-dependent friction force and a strong inertia-like behavior in the dynamics of driven particles.(26, 27) Theoretically, the inertia-like behavior of colloidal particles in viscoelastic fluids can be described equivalently by an effective mass of driven particles.(27) Moreover, it has been widely observed that the velocity of driven particles in viscoelastic micellar fluids is not steady but oscillate with time because of the nonlinear viscoelasticity.(28-30) In consistence, here, the velocity in acceleration does not increase monotonically but exhibit fluctuations and oscillation (Fig.1c and SFig.1a). To verify the oscillatory behavior of velocity, the motion is followed and tracked with a higher spatial resolution (0.25μm/pixel). It shows that the oscillatory behavior is an ambiguous dynamic characteristic in the acceleration (SFig.1b). Therefore, we suggest that the inertia-like acceleration and deceleration, to a large extent, results from the coupling between the motion of driven particles and the relaxation of deformation in a viscoelastic fluid, rather than from the mass of particles.

The deceleration time $\tau_d \sim$ 4ms means that as the frequency $f$ of SWE is below 125Hz, the velocity can completely relax to zero in the time of $E_g = 0$. As they are re-accelerated, a new direction of propulsion will be selected randomly by fluctuations such that velocities in different cycles are not correlated in direction. Therefore, Quincke rollers should behavior like Brownian particles at long time scales (>> T). This is supported by the observations at 10Hz and 40Hz where trajectories are characterized by random and sharp changes in direction (Fig.1d). The corresponding mean-square displacement (MSD) at long time scales (>>T) is approached by $t^1$ (10Hz and 40Hz in Fig.1e), a sign of Brownian motion. At short-time scales (< T), however, particles are propelled by a constant field $E_p$ and undergo ballistic motions in which MSD increases by $t^2$ as shown in Fig. 1e (10Hz and 40Hz).

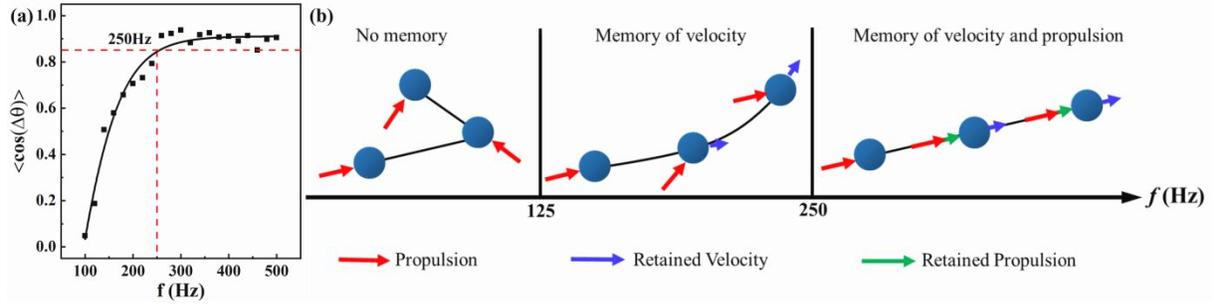

**Figure 2** Frequency-dependent dynamics at $E_p = 2.1E_c$ and $E_g = 0$ (a) Persistent index $\langle\cos(\Delta\theta)\rangle$ as a function of frequency. (b) Three regimes for memory. No memory occurs as the momentum is relaxed completely. Memory of velocity sets in as the momentum cannot be relaxed completely. As the charge relaxation is incomplete, a coupled memory of velocity and propulsion happens.

Above the frequency of 125Hz, the velocity cannot relax to zero before the particles are reaccelerated such that a memory of direction emerges. Correspondingly, the trajectory observed at 160Hz is much more directional than that of 10Hz and 40Hz. The corresponding MSD at long time scale (>T) increases by $t^2$, suggesting a long time ballistic motion. However, the trajectory at 160 Hz is distorted in direction at short time scale (Fig. 1d). The corresponding MSD is characterized by $t^{1.3}$ (< T) (Fig. 1e). It follows that at short time scale, the motion at 160Hz is still not well directed although a memory of velocity is present. To identify the exact frequency for persistent directional motion, a persistence index $\langle\cos(\Delta\theta)\rangle$(13) is calculated. Here, $\Delta\theta$ is the angle between velocities in two consecutive cycles. If velocities in different cycles have the same direction, $\langle\cos(\Delta\theta)\rangle = 1$. Figure 2a shows that at low frequencies, where Brownian motion dominates, $\langle\cos(\Delta\theta)\rangle$ is small. However, $\langle\cos(\Delta\theta)\rangle$ begins to increase with frequency (>100Hz) and reaches a plateau 0.90 as $f > 300$Hz. In this study, $\langle\cos(\Delta\theta)\rangle = 0.85$ is taken as the critical value for persistent directional motions. The corresponding frequency defines the critical frequency for persistent directional motion. Figure 2a shows that experimentally, persistent directional motion occurs only as $f > 250$Hz. This is much higher than the estimation (125Hz) based on $\tau_d$. This discrepancy indicates that there is another mechanism which affects the memory of direction as well as the deceleration.

Quincke rotation is driven by the torque $\vec{P} \times \vec{E}$. The direction of in-plane component $\vec{P}_\perp$ determines the direction of rotating and thus the direction of propulsion for translational motion. The propulsion for translational motion is antiparallel to $\vec{P}_\perp$(Fig. 1a). In equilibrium, the direction of translational velocity $\vec{v}$ is the same as that of translational propulsion. If the inertia of particle is negligible, the direction of $\vec{v}$ changes instantaneously with propulsion: they are coupled and parallel to each other in all time. However, Fig. 1c shows that the inertia-like behavior cannot be neglected anymore as $f > 125$Hz. In this case, if the relaxation time of $\vec{v}$ and the relaxation time of propulsion (charge) are not exactly the same, a deviation between the direction of velocity and the direction of propulsion becomes possible. The relaxation of propulsion is determined by the relaxation of polarization. If the relaxation time of $\vec{v}$, characterized by $\tau_d$, is shorter than the relaxation time of polarization, the direction of $\vec{v}$ is coupled with the direction of propulsion. No deviation between these two directions will occur. In this case, a transition from Brownian motion to persistent directional motion should occur at a frequency below 125Hz. However, this is in contradiction with the experimentally determined transition frequency 250Hz.

If the relaxation of $\vec{v}$ is slower than the relaxation of polarization, the memory of propulsion will emerge at a frequency $f_p$ above 125Hz. At the intermediate frequencies 125Hz $< f < f_p$, the direction of $\vec{v}$ is retained while the direction of propulsion will be re-selected in each cycle. A difference between the directions of propulsion and the direction of retained velocity gives rise to curved and partially directed trajectories as observed at 160Hz (Fig. 1d). Only as the frequency is above $f_p$ where the direction of propulsion and the direction of $\vec{v}$ are coupled and retained simultaneously, persistent directional motions become possible. This scenario is in consistence with our observations as shown in Fig.1d. The frequency 250Hz for persistent directional motions (Fig.2a) defines the critical frequency $f_p$ for the memory of propulsion. As a conclusion, the memory of velocity emerges at 125Hz whereas the memory of propulsion emerges at 250Hz. Between these two frequencies is a transition regime. Figure 2b summaries the frequency-dependent dynamics on the axis of frequency. The critical frequency 250Hz for persistent directional motion offers an estimation about the relaxation time of polarization: ~ 2ms. Theoretically, the relaxation time $\tau_{MW}$ of charge is determined by $\tau_{MW} = (\varepsilon_p + 2\varepsilon_l)/(\sigma_p + 2\sigma_l)$, which depends solely on the conductivities σ and permittivities ε of the particle (p) and liquid (l).(10) In Quincke system as used in this study, $\tau_{MW}$ was estimated as ~2.5ms.(13) This is in agreement with our result. Nevertheless, AOT is very hygroscopic and the conductivity of AOT solution is very sensitive to the mass ratio of AOT/water.(24) Therefore, the accurate value $\tau_{MW}$ is experimentally sensitive to the humidity of environment, which in practice is difficult to control. In this case, the observation on $\langle\cos(\Delta\theta)\rangle$ offers an alternative approach to determine $\tau_{MW}$.

The persistent directional motions at $f > 250$Hz is different from that driven by a constant (DC) field: the persistent directional motion under SWE is characterized by an oscillating speed (Fig. 3a). Most importantly, the maximum speed is far smaller than the value measured under a constant field of $E = E_p$. The mechanism is that to reach the equilibrium speed, a finite time $\tau_a$ for acceleration is necessary (Fig.1c). As the time of $E_p$ in one cycle is less than $\tau_a$ (~13ms, $f$ = 38Hz), Quincke particles cannot reach their equilibrium speed. Upon increasing frequency, the maximum speed in each cycle decreases as the time available for acceleration becomes less (Fig.3b). By contrast, the minimum speed is zero at $f$ < 125Hz but begins to increase with frequency as $f$ > 125Hz. The interpretation is that at $f$ <125Hz, the momentum completely relaxes to zero in the time of $E_g = 0$. At $f$ > 125Hz, however, particles are reaccelerated before the velocity drops to zero, leading to a non-zero minimum speed. The directional motion with oscillating speed can be approached by a ballistic motion with a mean speed at long time scale (>T), and the corresponding MSD increases by $t^2$ (Fig.1e, 300Hz). However, at short-time scale (< T), the oscillating of speed results in a damped MSD which increases by $t^{1.3}$ (160Hz and 300Hz, Fig. 1e).

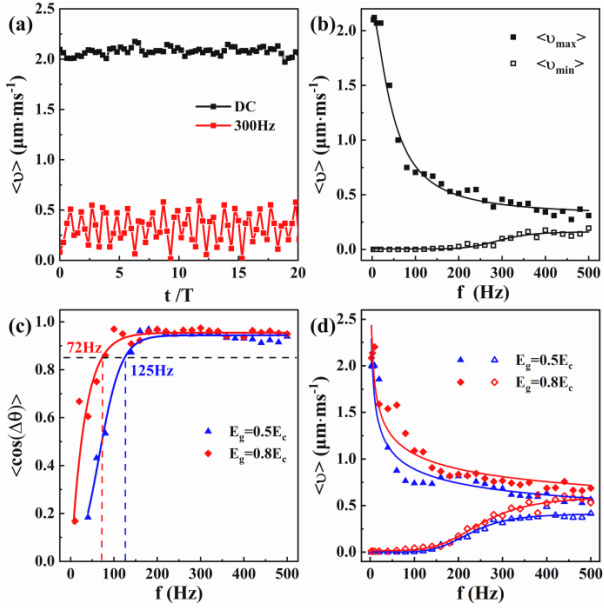

**Figure 3** Effect of $E_g$ on the dynamics. $E_p = 2.1E_c$. (a) Speeds of Quincke particles at direct current field and SWE. At 300Hz, the number of frames is experimentally not a constant in each cycle any more. (b) Mean speeds as a function of frequency at $E_g$=0. (c) Persistent index at $E_g = 0.5E_c$ and $E_g = 0.8E_c$. (d) Mean speeds as a function of frequency at positive $E_g$.

The relaxation time of momentum and therefore the critical frequency for the emergence of memory of velocity is experimentally determined by the deceleration time $\tau_d$, which is so far assumed as a constant and does not depend on the maximum speed before deceleration. This is supported by the observations that the deceleration time does not changes as well as the maximum speed decreases with frequency (>38Hz) (Sfig.1b). Additional evidence comes from the observations that the deceleration time are essentially the same even as the maximum speed is doubled by increasing $E_p$ from $1.7E_c$ to $2.7E_c$ (SFig.1c). These observations confirm that the deceleration is dominated by the physical properties of the viscoelastic fluid, which cannot be tuned temporally in one experiment. By contrast, the polarization is induced by the electric field. Therefore, the relaxation time of charge can be interfered by $E$. If a positive $E_g (< E_c)$ is applied, a finite torque is retained in the time of $E_g$. This torque is not strong enough to drive the particles to rotate but it can affect the relaxation of charge and then the memory of propulsion. To verify the effect of $E_g$, $\langle\cos(\Delta\theta)\rangle$ is measured at $E_g = 0.5E_c$ and $E_g = 0.8E_c$. The results show that the critical frequencies at $E_g = 0.5E_c$ and $E_g = 0.8E_c$ are 125Hz and 72Hz respectively (Fig. 3c). The corresponding relaxation times of charges are around 4.0ms and 7.0ms, which are remarkably longer than that at $E_g = 0$. Moreover, as $E_g$ increases, both the maximum and minimum speeds in each cycle become larger. However, the deceleration time measured at different $E_g$ are essentially the same as well (SFig. 1d). It follows that the deceleration process is dominated by physical properties of the fluid, and is decoupled with the relaxation of charge.

In passive colloids, as an electric field is applied, colloidal particles can aggregate to form crystalline structures because of the long-range attraction induced by electrohydrodynamic(EHD) flows.(31-35) Similarly, Quincke particles can also aggregate to form dense crystalline structures at subcritical electric fields $E < E_c$.(17) However, as Quincke rotation is activated, it is difficult for EHD attraction to confine the moving particles because of their high kinetic energy.(12, 19) However, the activity, in terms of speed, of Quincke particles under SWE can be tuned and reduced while they perform persistent directional motion at high frequencies (Fig.3a-b). It is expected that the low-activity particles may aggregate to form dense structures as well in the presence of the EHD attraction.

Figure 4a presents a dynamic dense cluster formed at $E_g = 0$ and 160Hz. The cluster exhibits a liquid-like structure and a flexible shape (Movie 1). The particles within the cluster run and turn in random directions periodically. This is distinct from the behavior of an isolated particle which exhibits a distorted directional motion (Fig. 1d). Moreover, the maximum speed of particles inside the clusters is significantly higher than that of isolated particles at 160Hz. The reason is that in the dense cluster, the dipolar repulsion between particles is strong and offers additional propulsion as $E_p$ is reset. Since the magnitude of repulsion is sensitive to local density and configuration, the instantaneous velocity of particles in clusters is not uniform (Fig. 4b). The critical frequency for persistent directional motion in dense clusters is 185Hz (Fig.4c), which is much lower than that isolated particles. This indicates that the relaxation of charge in dense clusters is much slower than that of isolated particles. The understanding is that the dipoles of particles produce a local electric field which can delay the relaxation of polarization as a positive $E_g$ does.

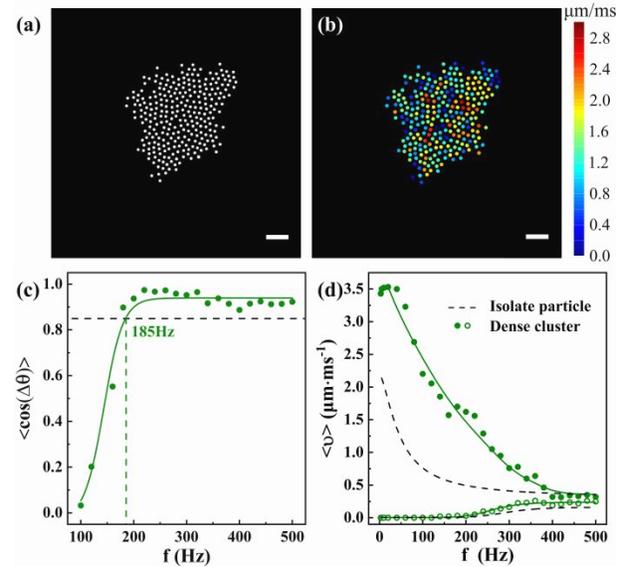

**Figure 4** Collective memory. $E_p = 2.1E_c$, $E_g = 0$ and $f$=160Hz. (a): Snapshot from experiment. (b) Particles are colored according to their speed. (c) Persistent index in dense clusters. (c) Persistent index as a function of frequency. (d) Mean speeds in dense clusters. Scale bar: 50 $\mu m$.

Similar dense clusters have been reported previously.(13) They interpreted their observations alone with the relaxation of polarization. The deviation between $\tau_{MW}$ and the experimentally identified time interval for directional motion was not explored. Meanwhile, the speed of Quincke particles therein was measured at large intervals of $E_p$, where Quincke particles can be fully accelerated, but they are taken as constant for all frequencies. Moreover, inertia-like behavior induced by the viscoelasticity of fluid and the EHD-induced attraction were not included to understand the formation of dynamic clusters in that study. Here, we show that to understand the complex behaviors of Quincke particles in a micellar fluid, it is critical to include the inertia-like behavior, charge relaxation and dipolar interaction.

## Conclusions

In summary, we show that the dynamic memory of Quincke particles arises from multiple mechanisms. Both the inertia-like behavior induced by the viscoelasticity and the relaxation of charge contribute to the memory. The memory and the activity are frequency-dependent, and can be tuned temporally by SWE. The density-dependent memory in dense clusters creates dynamic heterogeneity, which so far have not been included in the study of collective motions.

The dynamics of Quincke rollers so far has been assumed overdamped both experimentally and theoretically. However, the findings in this study reveal that in a micellar fluid, the viscoelasticity-induced memory experimentally can produce a strong inertia-like behavior in the dynamics of driven particles, and can significantly modify the dynamical characteristics of individuals and collective motions.

## Author Contributions

T.H.Z. designed the research. Y.Y. and M.F.Z. performed research and contributed equally. L.Z and T.H.Z. analyzed data and wrote the paper.

## Conflicts of interest

There are no conflicts to declare.

## Acknowledgements

T.H.Z. acknowledges financial support of the National Natural Science Foundation of China (Grant 11974255). L.Z. thanks Singapore Ministry of Education Academic Research Fund Tier 2 (MOE-T2EP50221-0012 & MOE-T2EP50122-0015).

## References

1. Ramaswamy S (2017) Active matter. *J. Stat. Mech. Theory Exp.* 2017(5):054002.
2. Vicsek T & Zafeiris A (2012) Collective motion. *Phys. Rep.* 517(3):71-140.
3. Shaebani MR, Wysocki A, Winkler RG, Gompper G, & Rieger H (2020) Computational models for active matter. *Nat. Rev. Phys.* 2(4):181-199.
4. Bialké J, Speck T, & Löwen H (2012) Crystallization in a Dense Suspension of Self-Propelled Particles. *Phys. Rev. Lett.* 108(16):168301.
5. Vicsek T, Czirók A, Ben-Jacob E, Cohen I, & Shochet O (1995) Novel Type of Phase Transition in a System of Self-Driven Particles. *Phys. Rev. Lett.* 75(6):1226-1229.
6. Chaté H (2020) Dry Aligning Dilute Active Matter. *Annu. Rev. Condens. Matter Phys.* 11(1):189-212.
7. Nagai KH, Sumino Y, Montagne R, Aranson IS, & Chaté H (2015) Collective Motion of Self-Propelled Particles with Memory. *Phys. Rev. Lett.* 114(16):168001.
8. Nejad MR, Doostmohammadi A, & Yeomans JM (2021) Memory effects, arches and polar defect ordering at the cross-over from wet to dry active nematics. *Soft Matter* 17(9):2500-2511.
9. Narinder N, Bechinger C, & Gomez-Solano JR (2018) Memory-Induced Transition from a Persistent Random Walk to Circular Motion for Achiral Microswimmers. *Phys. Rev. Lett.* 121(7):078003.
10. Jones TB (1984) Quincke Rotation of Spheres. *IEEE Transactions on Industry Applications* IA-20(4):845-849.
11. Das D & Saintillan D (2013) Electrohydrodynamic interaction of spherical particles under Quincke rotation. *Phys. Rev. E* 87(4):043014.
12. Bricard A, Caussin J-B, Desreumaux N, Dauchot O, & Bartolo D (2013) Emergence of macroscopic directed motion in populations of motile colloids. *Nature* 503(7474):95-98.
13. Karani H, Pradillo GE, & Vlahovska PM (2019) Tuning the Random Walk of Active Colloids: From Individual Run-and-Tumble to Dynamic Clustering. *Phys. Rev. Lett.* 123(20):208002.
14. Zhang B, Snezhko A, & Sokolov A (2022) Guiding Self-Assembly of Active Colloids by Temporal Modulation of Activity. *Phys. Rev. Lett.* 128(1):018004.
15. Yang Y, Zhang ZC, Qi F, & Zhang TH (2022) Emergence of Self-dual Patterns in Active Colloids with Periodical Feedback to Local Density. *arXiv:2204.07717 [cond-mat.soft]*.
16. Zhang B, Yuan H, Sokolov A, de la Cruz MO, & Snezhko A (2022) Polar state reversal in active fluids. *Nat. Phys.* 18(2):154-159.

17. Liu ZT, *et al.* (2021) Activity waves and freestanding vortices in populations of subcritical Quincke rollers. *Proc. Natl. Acad. Sci. USA* 118(40):e2104724118.
18. Zhang B, Karani H, Vlahovska PM, & Snezhko A (2021) Persistence length regulates emergent dynamics in active roller ensembles. *Soft Matter* 17(18):4818-4825.
19. Lu SQ, Zhang BY, Zhang ZC, Shi Y, & Zhang TH (2018) Pair aligning improved motility of Quincke rollers. *Soft Matter* 14(24):5092-5097.
20. Crocker JC & Grier DG (1996) Methods of Digital Video Microscopy for Colloidal Studies. *J. Colloid Interface Sci.* 179(1):298-310.
21. Han E, Zhu L, Shaevitz JW, & Stone HA (2021) Low-Reynolds-number, biflagellated Quincke swimmers with multiple forms of motion. *Proc. Natl. Acad. Sci. USA* 118(29):e2022000118.
22. Pradillo GE, Karani H, & Vlahovska PM (2019) Quincke rotor dynamics in confinement: rolling and hovering. *Soft Matter* 15(32):6564-6570.
23. Mukherjee K, Moulik SP, & Mukherjee DC (1993) Thermodynamics of micellization of Aerosol OT in polar and nonpolar solvents. A calorimetric study. *Langmuir* 9(7):1727-1730.
24. Hsu MF, Dufresne ER, & Weitz DA (2005) Charge Stabilization in Nonpolar Solvents. *Langmuir* 21(11):4881-4887.
25. Gomez-Solano JR & Bechinger C (2014) Probing linear and nonlinear microrheology of viscoelastic fluids. *EPL* 108(5):54008.
26. Goychuk I (2022) Memory can induce oscillations of microparticles in nonlinear viscoelastic media and cause a giant enhancement of driven diffusion. *Proc. Natl. Acad. Sci. USA* 119(48):e2205637119.
27. Berner J, Müller B, Gomez-Solano JR, Krüger M, & Bechinger C (2018) Oscillating modes of driven colloids in overdamped systems. *Nat. Commun.* 9(1):999.
28. Jayaraman A & Belmonte A (2003) Oscillations of a solid sphere falling through a wormlike micellar fluid. *Phys. Rev. E* 67(6):065301.
29. Zhang Y & Muller SJ (2018) Unsteady sedimentation of a sphere in wormlike micellar fluids. *Phys. Rev. Fluids* 3(4):043301.
30. Mrokowska MM & Krztoń-Maziopa A (2019) Viscoelastic and shear-thinning effects of aqueous exopolymer solution on disk and sphere settling. *Sci. Rep.* 9(1):7897.
31. Zhang TH & Liu XY (2014) Experimental modelling of single-particle dynamic processes in crystallization by controlled colloidal assembly. *Chem. Soc. Rev.* 43(7):2324-2347.
32. Mittal M, Lele PP, Kaler EW, & Furst EM (2008) Polarization and interactions of colloidal particles in ac electric fields. *J. Chem. Phys.* 129(6):064513.
33. Ristenpart WD, Aksay IA, & Saville DA (2007) Electrically Driven Flow near a Colloidal Particle Close to an Electrode with a Faradaic Current. *Langmuir* 23(7):4071-4080.
34. Hong Yu Chen LW & Zhang TH (2021) Symmetry-Dependent Kinetics of Dislocation Reaction. *Chin. Phys. Lett.* 38(6):066101.
35. Zhang TH, Zhang ZC, Cao JS, & Liu XY (2019) Can the pathway of stepwise nucleation be predicted and controlled? *Phys. Chem. Chem. Phys.* 21(14):7398-7405.

# Supplementary Information

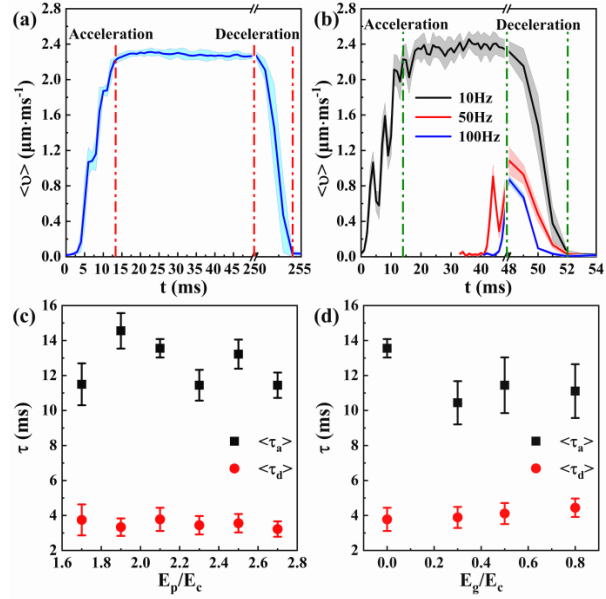

SFigure 1 (a) Evolution of speed in one cycle. The speed is averaged over thousands of particles. (b) Acceleration and deceleration observed at different frequencies with a resolution of 0.25μm/pixel. The error bars in (a) and (b) are represented by shade. (c) Effect of the peak value $E_p$ on the acceleration time and deceleration time. (d) Effect of the ground value on the acceleration time and deceleration time. In (a), (b) and (d), $E_p = 2.1E_c$.

**Description of movies**

Movie 1: Dynamic Cluster. $E_p = 2.1E_c$, $E_g = 0$ and $f = 160Hz$. The movie runs for $100\ ms$ of real time. The field of view is $590 \times 590 \mu m^2$. Scale bar: $50\ \mu m$.